\documentclass[aps,pra,twocolumn,superscriptaddress,showpacs]{revtex4}
\usepackage{graphicx}
\usepackage{simon}
\usepackage{microtype}

\begin{document}

\title{Orientation-to-alignment conversion and spin squeezing}

\author{S.~M.~Rochester}
\homepage{http://rochesterscientific.com/}
\affiliation{Rochester Scientific, El Cerrito, California,
94530-1757}
 \affiliation{Department of Physics, University of California at
Berkeley, Berkeley, California 94720-7300}

\author{M.~P.~Ledbetter}
\affiliation{Department of Physics, University of California at Berkeley, Berkeley, California 94720-7300}

\author{T.~Zigdon}
\affiliation{Department of Chemistry, Bar-Ilan University, Ramat Gan 52900, Israel}

\author{A.~D.~Wilson-Gordon}
\affiliation{Department of Chemistry, Bar-Ilan University, Ramat Gan 52900, Israel}

\author{D. Budker}
 \email{budker@berkeley.edu}
 \affiliation{Department of Physics, University of California at
Berkeley, Berkeley, California 94720-7300}
 \affiliation{Nuclear Science Division, Lawrence
 Berkeley National Laboratory, Berkeley, California 94720}

\date{\today}

\begin{abstract}
The relationship between orientation-to-alignment conversion (a
form of atomic polarization evolution induced by an electric
field) and the phenomenon of spin squeezing is demonstrated. A
``stretched'' state of an atom or molecule with maximum
angular-momentum projection along the quantization axis
possesses orientation and is a quantum-mechanical
minimum-uncertainty state, where the product of the equal
uncertainties of the angular-momentum projections on two
orthogonal directions transverse to the quantization axis is
the minimum allowed by the uncertainty relation. Application of
an electric field for a short time induces
orientation-to-alignment conversion and produces a
spin-squeezed state, in which the quantum state essentially
remains a minimum-uncertainty state, but the uncertainties of
the angular-momentum projections on the orthogonal directions
are unequal. This property can be visualized using the
angular-momentum probability surfaces, where the radius of the
surface is given by the probability of measuring the maximum
angular-momentum projection in that direction. Brief remarks
are also given concerning collective-spin squeezing and quantum
nondemolition measurements.
\end{abstract}


\maketitle

\section{Introduction}

Since the pioneering work of Kitagawa and Ueda \cite{Masahito},
the concept of spin squeezing, or the redistribution of
uncertainties from one spin component to another, has drawn
significant attention \cite{Ma2010}. Reducing the uncertainty
in a particular spin component to be measured at the expense of
others can, in principle, allow measurements at the fundamental
Heisenberg limit of uncertainty, which scales as $1/F$ for the
relative uncertainty of a measurement of the projection of an
effective angular momentum $F$, rather than at the standard
quantum limit, which scales as $1/\sqrt{F}$.

One area in which spin squeezing is of practical interest is
optical magnetometry---the idea of gaining sensitivity via spin
squeezing is an attractive one. Unfortunately, this application
is not as straightforward as it may seem, and in fact, there is
no sensitivity gain in a rather broad class of situations
\cite{Auzinsh2004QND}. Nevertheless, as has been shown
recently, squeezing can lead to increase in bandwidth
\cite{Shah2010,Wasilewski2010}, and can increase sensitivity in
cases involving nonexponential relaxation \cite{Vasilakis2011}.

In this note, we discuss two aspects of spin squeezing relevant
to atomic magnetometry. The first concerns the relationship
between spin squeezing and a type of polarization evolution
known as alignment-to-orientation conversion (AOC), which
occurs when an electric field is applied to a polarized atomic
ensemble. Here ``orientation'' refers to the rank-one atomic
polarization moment having a preferred direction, and
``alignment'' to the rank-two polarization moment with a
preferred axis but no preferred direction.
Alignment-to-orientation conversion is an important mechanism
for atomic magnetometry, occurring, for example, in nonlinear
Faraday rotation \cite{Bud2000AOC}. It also occurs in other
areas such as nuclear quadrupole resonance (NQR)
\cite{Bud2003NQR}, and has been extensively studied for many
years.

Here we point out that there is a close relationship between
AOC and spin squeezing: when an atom in a stretched state is
placed in an orthogonal electric field, spin squeezing is
caused as a result of, in this case, the inverse process of
orientation-to-alignment conversion (OAC). We quantify the
amount of squeezing that can be obtained, and illustrate the
process using a polarization visualization technique. The
electric field needed to produce the squeezing can be either dc
or off-resonant ac. In fact, the latter has already been used
to generate spin squeezing in the ground state of alkali atoms
(see Refs.\ \cite{Chaudhury2007,Fernholz2008}; this case is
analyzed in Appendix C).

Note that there is an essential difference between the
squeezing produced by AOC and that discussed in Ref.\
\cite{Masahito}: in the latter case the squeezing is produced
by an operator acting on the collective spin of the ensemble,
while the electric-field Hamiltonian that induces AOC acts on
the individual spin of each atom. From a practical standpoint,
it is much more desirable to squeeze the collective spin,
rather than the individual spins, because the effective angular
momentum participating in the scaling discussed above can be
made very large.

We then discuss a second aspect of spin squeezing that has been
demonstrated in the context of optical
magnetometry---collective spin squeezing via a QND interaction.
Here we remark on the origin of increased noise in the
unobserved spin quadrature that ensures compliance with the
uncertainty relations for orthogonal spin projections.

\section{Spin squeezing by interaction with the electric field}
\label{subsec:equation}

Measurements involving quantum systems are fundamentally
limited by uncertainty relations derived from the commutation
relations of quantum-mechanical operators. A textbook example
is a state of a system (such as an atom or a molecule; we will
henceforward refer to an atom) with total angular momentum $F$
prepared in a ``stretched'' state with a fixed projection $m$
on a chosen quantization axis ($z$) such that $|m|=F$, i.e.,
the state $\ket{F,m=F}$. A state stretched along another
direction can be written by rotating $\ket{F,F}$ using the
quantum-mechanical rotation operator. For the state
$\ket{F,F}_{\uv{x}}$ stretched along $\uv{x}$ this gives
\cite{Masahito}
\begin{equation}
\ket{F,F}_\uv{x} = 2^{-F}\sum^{2F}_{k=0}
  \binom{2F}{k}^{\frac{1}{2}}\ket{F,F-k}, \label{Eq:statecoupled}
\end{equation}
where the binomial coefficients are given by
\begin{equation}
    \binom{n}{k}=\frac{n!}{k!(n-k)!}.
\end{equation}
If an experiment is performed that measures the $x$-projection
of $\ket{F,F}_{\uv{x}}$, the outcome is always the same
($F_x=F\hbar$). On the other hand, measuring the projection on
an orthogonal axis, say $y$, one measures zero on average,
$\abr{F_y}=0$, but each specific measurement can yield any
result such that $-F\le F_y/\hbar \le F$, with a similar result
for the projection on $z$. The uncertainty relation for the
angular-momentum projections reads:
\begin{equation}
\Delta F_y \Delta F_z \ge \frac{\hbar^2}{2}F,\label{Eq:UncertCond}
\end{equation}
where the uncertainty in $\Delta F_y$ is defined according to
\begin{equation}
\Delta F_y = \sqrt{\abr{F_y^2}-\abr{F_y}^2},
\end{equation}
and similarly for $\Delta F_z$. Explicitly calculating the
uncertainties $\Delta F_y$ and $\Delta F_z$ using the
appropriate quantum-mechanical operators, we find that, as
expected from symmetry, these uncertainties are equal, and that
their values realize the equality in the expression
\eqref{Eq:UncertCond}, which means that the stretched state is
a \emph{minimum-uncertainty state}.

\begin{figure}
\includegraphics{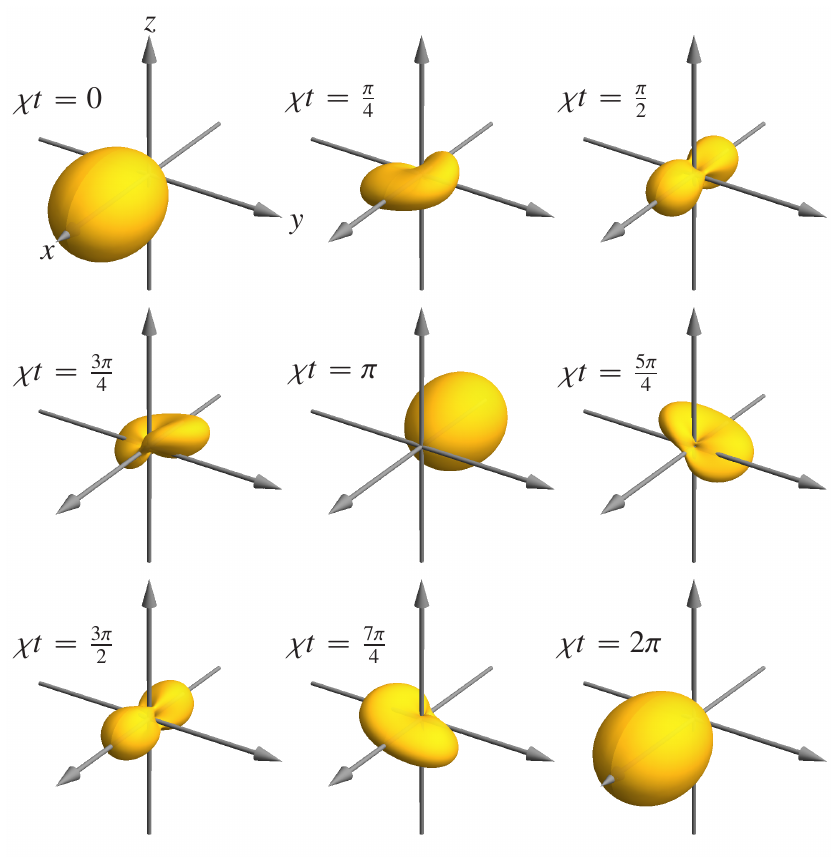}
\caption{(Color online.) Angular-momentum probability surfaces (see text)
showing an $F=2$ atomic state initially stretched along
$\uv{x}$ and evolving in the presence of an electric field
along $\uv{z}$. A complete cycle of orientation-to-alignment
conversion is shown.}\label{fig:LongAMPS}
\end{figure}

One way of visualizing the state is using angular-momentum
probability surfaces (AMPS) \cite{Auz97,Roc2001,Ale2005}. The
radius of the surface in a given direction is proportional to
the probability to measure the maximum angular-momentum
projection ($=F$) in that direction. This corresponds to the
quasi-probability distribution plotted in Ref.\ \cite{Masahito}
and is the analog, for spin states, of the Q-function of
quantum optics. The AMPS for $\ket{F,F}_{\uv{x}}$ is shown in
the upper-left plot of Fig.\ \ref{fig:LongAMPS}. It is clearly
pointing in the $x$ direction, and is symmetric about $\uv{x}$.

We next consider the evolution of the stretched state under the
influence of an electric field. Let us assume that the field is
applied along the quantization axis $z$. For simplicity, we
assume that the electric field is either dc or linearly
polarized off-resonant ac (the two cases are essentially
equivalent \cite{Chaudhury2007,Fernholz2008}). The Hamiltonian
of the system in the presence of the electric field is
\begin{equation}
H_E = -\frac{1}{2}\alpha_0E^2-\frac{1}{2}\alpha_2E_z^2\frac{3F_z^2-\V{F}^2}{\hbar^2 F(2F-1)}.
\end{equation}
where $\alpha_0$ is the scalar polarizability of the state, and
$\alpha_2$ is the tensor polarizability. (The vector
polarizability does not contribute to the Hamiltonian, as we
are considering a linearly polarized electric field.) We
neglect the scalar polarizability term and the part of the
tensor polarizability that is independent of $F_z$, since they
result only in a common shift of the Zeeman sublevels of the
ground or excited hyperfine state, and so do not lead to
evolution of the Zeeman polarization. We therefore consider the
following Hamiltonian for each spin in the ensemble:
\begin{equation}
H_E = \chi F_z^2/\hbar,\label{Eq:IntHamiltonian}
\end{equation}
where we have defined
\begin{equation}
    \chi=-\frac{3}{2\hbar F(2F-1)}\alpha_2E_z^2.
\end{equation}
This Hamiltonian has the same form as the one presented by
Kitagawa and Ueda \cite{Masahito} for ``one-axis twisting.''
However, there is an essential difference in that $F_z$ in
Ref.\ \cite{Masahito} refers to the collective spin of an
ensemble of particles, whereas we are considering a single atom
or, equivalently, an ensemble of uncorrelated atoms. The
Hamiltonian $H_E$ generates the unitary transformation
\begin{equation}
U(t) = e^{-i\chi t F^{2}_z/\hbar^2}. \label{Eq:Utransformation}
\end{equation}

The evolution of $\ket{F,F}_{\uv{x}}$ in the Schr\"odinger
picture according to the evolution operator $U(t)$ is
illustrated in Fig.\ \ref{fig:LongAMPS} for the case of $F=2$.
The state oscillates between having a preferred direction,
indicating that it possesses orientation (for example, at $\chi
t=0$), and having a preferred axis, indicating that it
possesses alignment (for example, at $\chi t=\pi/2$); we
therefore refer to this kind of evolution as OAC or AOC.

We note that the evolution of polarized atomic and molecular
states in the presence of an electric field has been
extensively studied in the literature (see, for example,
\cite{Auz2006,Auz97,Ale2005} and references therein). However,
it has not been broadly recognized that spin squeezing is
naturally associated with this evolution. In fact, we will now
see that, as the process of OAC begins, the uncertainty of a
spin measurement along a particular axis perpendicular to $x$
is reduced, while that along the orthogonal axis is increased,
so that the state, to first order, remains a minimum
uncertainty state. To show this, we explicitly calculate the
uncertainty in the measurement of $F_y$ as a function of time;
a rotation about the $x$-axis allows us to find the axis of
optimal squeezing.

Following Ref.\ \cite{Masahito}, we analyze the means and
variances of the operators $F_x(t), F_y(t)$ in the Heisenberg
picture using the raising and lowering operators
\begin{equation}
F_\pm (t) = F_x(t) \pm i F_y(t), \label{Eq:lowupoperators}
\end{equation}
which evolve in the Heisenberg picture as
\begin{equation}
\begin{split}
F_+ (t) &= U(t)^{\dag}F_+ (0)U(t)\\
&=e^{i\chi t F^{2}_z/\hbar^2}F_+ (0)e^{-i\chi t F^{2}_z/\hbar^2}\\
&=F_+ (0)e^{2i\chi t(F_z/\hbar+\frac{1}{2})} \label{Eq:upoperator},
\end{split}
\end{equation}
\begin{equation}
F_- (t) = [F_+ (t)]^{\dag}=e^{-2i\chi t(F_z/\hbar+\frac{1}{2})}F_-(0) \label{Eq:downoperator},
\end{equation}
where $F_-(0)=F^{\dag}_+(0)$. The details of the derivation of
Eqs.\ (\ref{Eq:upoperator},\ref{Eq:downoperator}) are presented
in Appendix A. The components of $F_x(t), F_y(t)$ are now given
by:
\begin{equation}
\begin{split}
    F_x (t)
    &=\frac{1}{2}[F_+(t) + F_-(t)]\\
    &=\frac{1}{2}[F_+(0)e^{2i\chi t(F_z/\hbar + \frac{1}{2})}
    + e^{-2i\chi t(F_z/\hbar + \frac{1}{2})}F_-(0)], \label{Eq:Fx}
\end{split}
\end{equation}
\begin{equation}
\begin{split}
    F_y (t)
    &=\frac{1}{2i}[F_+(t)-F_-(t)]\\
    &=\frac{1}{2}[F_+(0)e^{2i\chi t(F_z/\hbar+\frac{1}{2})}
    -e^{-2i\chi t(F_z/\hbar+\frac{1}{2})}F_-(0)]. \label{Eq:Fy}
\end{split}
\end{equation}
In Appendix B, we present a detailed calculation of the
expectation value $\abr{F_x}$ [Eq.\ \eqref{Eq:meanvalFx}]. The
means and variances of the other components of the angular
momentum can be calculated in a similar manner. In order to
analyze the mean values and variances of the spin projections
along all directions transverse to $x$, it is convenient to
write $F_y(t)$ in a coordinate frame obtained from the original
one by rotating about $\uv{x}$ by an angle $\nu$, according to
the unitary transformation
\begin{equation}
F_{y,\nu}=e^{i\nu F_x (t)/\hbar}F_ye^{-i\nu F_x (t)/\hbar}.
\end{equation}
The expectation values of the components of the angular
momentum become
\begin{subequations}
\begin{equation}
\langle F_x \rangle = \hbar F (\cos\chi t)^{2F-1},\label{Eq:meanvalFx}
\end{equation}
\begin{equation} \langle F_{y,\nu} \rangle = 0,\label{Eq:meanvalFy}
\end{equation}
 \begin{equation}\langle F_{z,\nu} \rangle = 0,\label{Eq:meanvalFz}
 \end{equation}
\end{subequations}
while the variances are given by
\begin{subequations}\label{Eq:variancesF}
\begin{equation}
(\Delta F_x)^2 = \hbar^2\frac{F}{2}[2F(1-\cos^{2(2F-1)} \chi t)-(F-\frac{1}{2})A],\label{Eq:variancesFx}
\end{equation}
\begin{equation}
(\Delta F_{y,\nu})^2 = \hbar^2\frac{F}{2}\{1+\frac{1}{2}(F-\frac{1}{2})[A+\sqrt{A^{2}+B^{2}}\cos(2\nu+2\delta)]\},\label{Eq:variancesFy}
\end{equation}
\begin{equation}
(\Delta F_{z,\nu})^2 = \hbar^2\frac{F}{2}\{1+\frac{1}{2}(F-\frac{1}{2})[A-\sqrt{A^{2}+B^{2}}\cos(2\nu+2\delta)]\},\label{Eq:variancesFz}
\end{equation}
\end{subequations}
where $A = 1- \cos^{2F-2} 2\chi t$, $B = 4\sin \chi t
\cos^{2F-2} \chi t,$ and
$\delta=\frac{1}{2}\arctan\frac{B}{A}$. According to Eqs.\
\eqref{Eq:variancesF}, $\Delta F_{y,\nu}$ is minimized and
$\Delta F_{z,\nu}$ is maximized when $\cos(2\nu+2\delta)=-1$,
i.e., $\nu=\frac{\pi}{2}-\delta$. This determines the axis with
the best squeezing at a given time.

\begin{figure}
\includegraphics{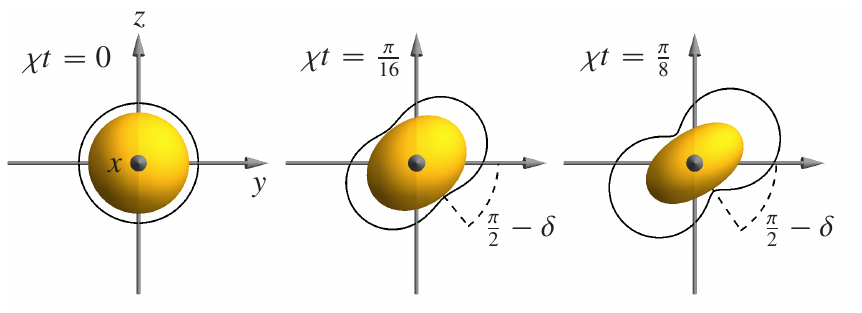}
\caption{(Color online.) The initial part of the evolution shown in Fig.\
\ref{fig:LongAMPS} as viewed from the $x$ direction, showing
the process of squeezing. The solid line is a polar plot of
$\Delta F_{y,\nu}$ as a function of azimuth, with the angle
$\nu=\pi/2-\delta$ of the minimum-uncertainty axis indicated by
the dashed line.}\label{fig:ShortAMPS}
\end{figure}

Considering the evolution shortly after the application of the
electric field, we find that, to first order, the state remains
a minimum-uncertainty state. However the uncertainties in $F_y$
and $F_z$ are no longer equal [Eqs.\ \eqref{Eq:variancesF}];
therefore, we have generated a \emph{spin-squeezed state}
(SSS). This is indicated by AMPS plots observed along the
$x$-axis (Fig.\ \ref{fig:ShortAMPS}). The utility of SSS is
that, in principle, they allow an improved sensitivity in
certain appropriately designed measurements. [The fundamental
quantum-mechanical limit on the uncertainty of a measurement of
$F_x$ is not $\hbar \sqrt{F/2}$ as implied by the uncertainty
relation \eqref{Eq:UncertCond} for the case of $\Delta
F_x=\Delta F_y$ (the standard quantum limit, or SQL), but the
Heisenberg limit \cite{BollingerPRA541996,PhysRevLett.86.4431}
of $\hbar/\sqrt{2}$.]

\begin{figure}
\includegraphics{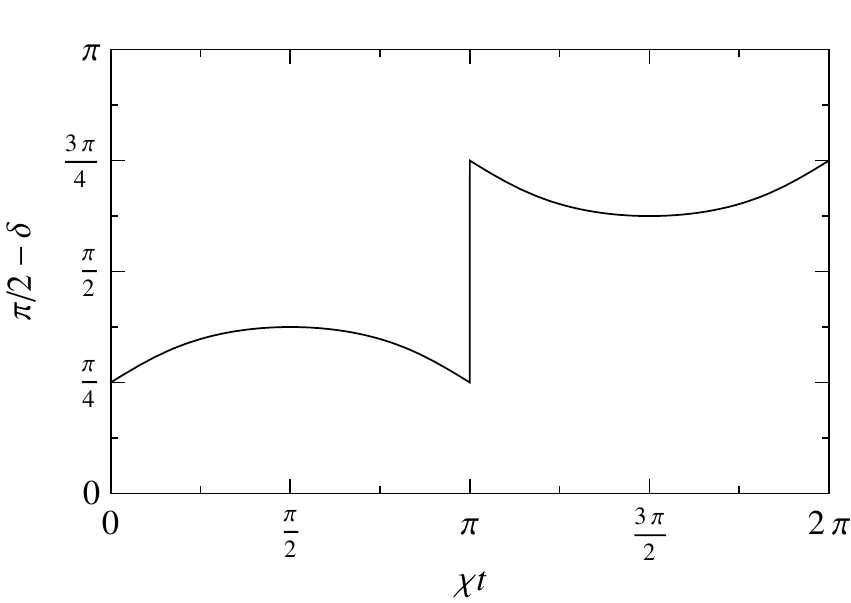}
\caption{The angle $\nu=\frac{\pi}{2}-\delta$ of the minimum
uncertainty axis as a function of time for a state with $F=2$.
The $\pi/2$ shift halfway through the cycle corresponds to the
mirror symmetry seen between the first and second halves of the
cycle in Fig.\ \ref{fig:LongAMPS}.}\label{fig:DeltaTimeDependence}
\end{figure}

Wineland \textit{et al.}\ \cite{WinelandPRA461994} defined a
squeezing parameter to indicate sensitivity to rotations of the
angular-momentum states. Considering squeezing along the $y$
axis rotated by an angle $\nu$ about $\uv{x}$, the squeezing
parameter $\xi_{R}$ is the uncertainty in the rotation of the
spin, $\Delta F_{y,\nu}/\abs{\abr{F_x}}$, normalized to the
uncertainty $1/\sqrt{2F}$ expected in the SQL, i.e., the
uncertainty obtained using a stretched state:
\begin{equation}\label{Eq:squeezingparameter}
\xi_{R} = \sqrt{2F}\frac{\Delta F_{y,\nu}}{\abs{\abr{F_x}}}.
\end{equation}
Substituting from Eqs.\ \eqref{Eq:meanvalFx} and
\eqref{Eq:variancesFy}, we find the squeezing along the
minimum-uncertainty axis to be
\begin{equation}\label{Eq:minsqueezing}
\xi_{R} = \frac{\sqrt{1+\frac{1}{2}\prn{F-\frac{1}{2}}\prn{A-\sqrt{A^{2}+B^{2}}}}}
{\abs{(\cos\chi t)^{2F-1}}}.
\end{equation}

\begin{figure}
  \includegraphics{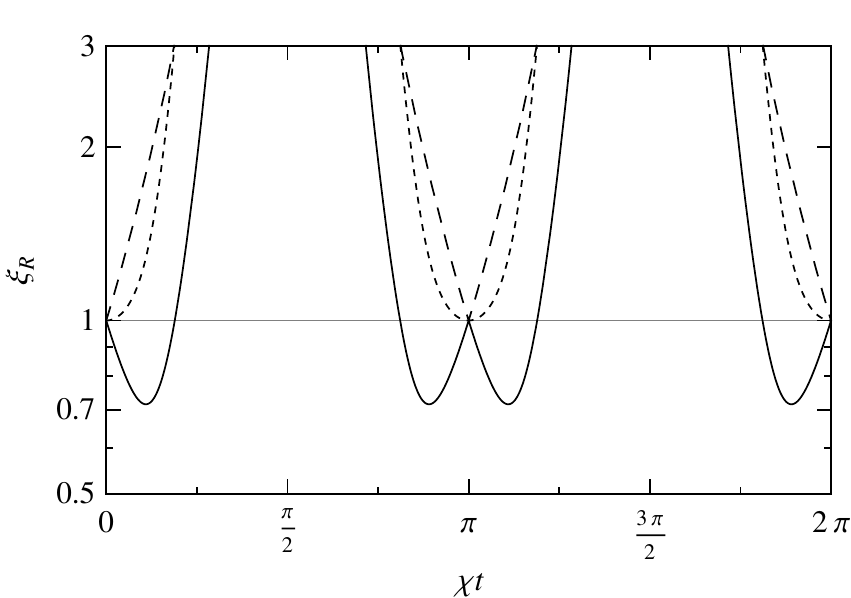}
  \caption{The squeezing parameter $\xi_{R}$ for $F=2$ as a
  function of time. The solid curve shows $\xi_{R}$ for the
  most strongly squeezed axis ($\nu = \frac{\pi}{2}-\delta$),
  while the dashed curve gives $\xi_{R}$ for the orthogonal
  ``anti-squeezing'' axis. A value $\xi_{R}<1$ (below the gray line)
  indicates a squeezed state. The dotted curve shows the
  product of the squeezing parameters for the two orthogonal
  axes, with a value of 1 indicating a minimum-uncertainty
  state.}\label{fig:SqueezingTimeDependence}
\end{figure}

In Fig.\ \ref{fig:SqueezingTimeDependence} we plot the
squeezing parameter for the minimum-uncertainty axis as a
function of time for $F=2$. We observe that $\xi_{R}$ initially
decreases below unity, indicating a squeezed state, but
subsequently tends toward infinity at $t=\frac{\pi}{2}+k\pi$,
with integer $k$, because $\abs{\abr{F_x}}$ tends to zero. From
Eq.\ \eqref{Eq:meanvalFx} we see that this is true for any
value of $F$. If the electric field inducing OAC is turned off
at the time corresponding to the minimum value of $\xi_{R}$, we
obtain squeezing along a fixed axis whose direction can be
found from Fig.\ \ref{fig:DeltaTimeDependence}. (Note that in
some cases it is desirable to have the squeezing axis rotate in
time---this could be accomplished here by subsequent
application of a magnetic field.)

\begin{figure}
  \includegraphics{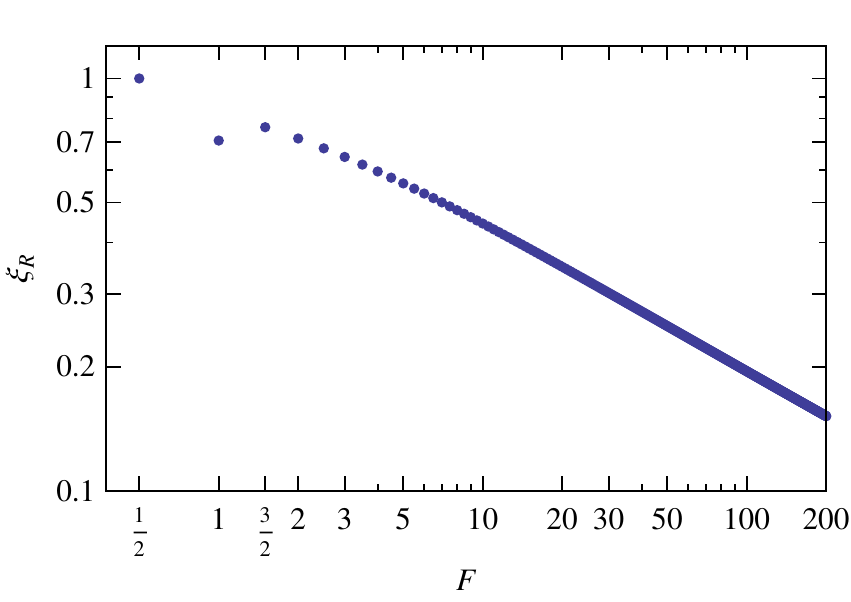}
  \caption{(Color online.) Logarithmic plot of the minimum
  value of $\xi_{R}$ with respect to $t$ as a function of $F$
  when $\nu = \frac{\pi}{2}-\delta$. The dependence quickly
  approaches a power law ($\xi_{R}\propto F^{-1/3}$).
  Note that the case of $F=1$ has an apparently anomalously
  small value of $\xi_{R}$. This is a special case in which the
  optimum squeezing parameter is achieved as $\chi t$
  approaches $\pi/2$, when the spin projection $\abr{F_x}$ goes
  to zero.}\label{fig:SqueezingFDpendence}
\end{figure}

In Fig.\ \ref{fig:SqueezingFDpendence} we plot the minimum of
$\xi_{R}$ with respect to $t$ as a function of $F$. This plot
differs from the corresponding plot in Ref.\ \cite{Masahito}
because we use a different definition of the squeezing
parameter. To find the asymptotic behavior of $\xi_{R}$ for
large $F$, we note that as $F$ increases, the time $\Gc t$ at
which the squeezing is minimized decreases faster than
$1/\sqrt{F}$, but slower than $1/F$. Thus, for $F\gg1$, we can
assume that, near the minimum, the parameters $\Gg=1/(\Gc tF)$
and $\Gb=\Gc^2t^2F$ are both small. Writing the square of the
squeezing parameter \eqref{Eq:minsqueezing} in terms of $\Gg$
and $\Gb$, we expand to second order and find
\begin{equation}\label{Eq:minsqueezingbg}
\xi_{R}^2 \approx \frac{2\Gb^2}{3}+\frac{\Gg^2}{4}.
\end{equation}
Substituting back for $F$ and $t$, we minimize with respect to
$t$ and find that at the time
\begin{equation}
\Gc t_\text{min}\approx\frac{3^{1/6}}{2^{2/3}}F^{-2/3}
\end{equation}
the minimum squeezing parameter is given by
\begin{equation}
\xi_{R}^\text{min}\approx\frac{3^{1/3}}{2^{5/6}}F^{-1/3},
\end{equation}
equivalent to the result found in Ref.\ \cite{Sorensen2001}.
Relatively large values of $F$ available for single atoms are
$F=4$ in the ground state of Cs (see Appendix C for a
calculation of squeezing in this system) and $F=12.5$ in a
metastable state of Dy \cite{Bud94Dy}. Even higher values of
$F$ are attainable in Rydberg atoms and in molecules with large
rotational excitation. Very large effective values of $F$ can
be achieved in a somewhat different situation in which
squeezing is done on a correlated ensemble, as discussed below.

\begin{figure}
  \includegraphics{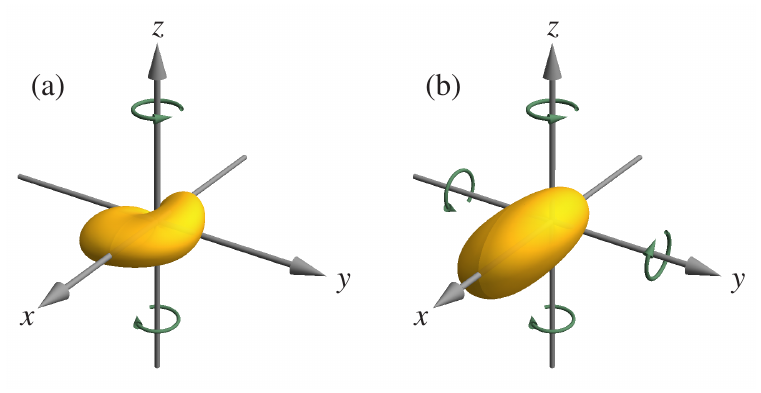}
  \caption{(Color online.) Comparison of AMPS for the (a)
  ``single-axis twisting'' and (b) ``two-axis countertwisting''
  Hamiltonians at times equal to $1/8$ of the respective
  quantum-beat periods for an $F=2$ state initially stretched along
  $\uv{x}$. Twisting about the $z$-axis introduces
  ``swirliness'' in (a); simultaneous twisting about the
  $y$-axis (b) cancels this effect.}\label{fig:OneAxisTwoAxis}
\end{figure}

The polarization evolution due to the Hamiltonian
\eqref{Eq:IntHamiltonian} is termed ``single-axis twisting'' by
Kitagawa and Ueda \cite{Masahito} because the effect on the
angular-momentum probability distribution can be visualized as
resulting from a twisting motion about the $z$-axis (Fig.\
\ref{fig:OneAxisTwoAxis}a). For the purposes of generating
squeezed states, this type of evolution has some drawbacks
resulting from the asymmetry of the evolution with respect to
the $z$- and $y$-axes. First, as we have seen in Fig.\
\ref{fig:DeltaTimeDependence}, the optimal squeezing axis
changes as a function of time. Second, the distortion in the
probability distribution introduced by the twisting motion,
described as ``swirliness'' in Ref.\ \cite{Masahito}, limits
the maximum squeezing that can be obtained. These effects can
be obviated by creating a more symmetric Hamiltonian in which
twisting is performed about two orthogonal axes (``two-axis
countertwisting''), as shown in Fig.\
\ref{fig:OneAxisTwoAxis}(b). Here we have plotted the effect of
a Hamiltonian of the form $H=\chi(F_z^2-F_y^2)/\hbar$.
Interaction described by such a Hamiltonian can be achieved by
the use of two incoherent fields, such as a light field along
with a magnetic field inducing the nonlinear Zeeman effect
\cite{Fernholz2008}, two light fields of different frequencies,
or a light field and a static electric field (see Ref.\
\cite{Ma2010} and references therein). This removes the
swirliness, fixes the squeezing axis, and allows the maximum
amount of squeezing to be attained. We do not further consider
the two-axis Hamiltonian here, as the simpler Hamiltonian
\eqref{Eq:IntHamiltonian} illustrates the principle under
discussion.

The value for spin squeezing found here differs from that
reported by Kitagawa and Ueda \cite{Masahito} because we use a
form of the squeezing parameter, normalized to the expectation
value of $F_x$, that is appropriate for measurements of
rotation. There is a second, more fundamental difference
between the two results, however, stemming from the definition
of the interaction Hamiltonian. While Kitagawa and Ueda's
\cite{Masahito} interaction Hamiltonian relates to an ensemble
with quantum correlations among the individual spins
\cite{WinelandPRA461992}, our $H_E$ \eqref{Eq:IntHamiltonian}
operates on each individual spin in the ensemble without
creating correlation. This difference shows up clearly in the
case in which all the atoms are in a $F=1/2$ state. For an
individual $F=1/2$ atom, there is no differential shift induced
by $H_E$, and, in any case, the only possible polarized state
is a stretched state (the highest-rank polarization possible is
orientation). Thus no squeezing is possible in this case. (For
the common experimental case of alkali atoms, with electronic
spin $J=1/2$, the additional nuclear spin---resulting in higher
total angular momentum---allows spin squeezing
\cite{Fernholz2008}.) However, for $N$ spin 1/2 atoms in the
case in which the Hamiltonian acts on the collective spin
$F_z^{(\text{tot.})}=\sum_i F_z^{(i)}$, the correlation between
the particles can result in squeezing as for a fictitious
particle with $F=N/2$. This means that very large effective
values of $F$ can be obtained, which would not be feasible for
an uncorrelated ensemble for which $F$ is the angular momentum
of each polarized atom. Some brief remarks on methods of
producing squeezing of a correlated ensemble are given in the
next section.

\section{Remarks on ensemble squeezing}

Various techniques have been introduced in recent years to
achieve collective-spin squeezing in atomic ensembles. These
include methods based on collective interactions in
Bose-Einstein condensates (see Ref.\ \cite{Riedel2010} and
references therein), employing collective interactions between
atoms via an optical cavity (see Ref.\ \cite{SchleierSmith2010}
and references therein), or methods based on generating
squeezing by performing so-called quantum nondemolition (QND)
measurements on an ensemble
\cite{Appel2009,Takano2009,SchleierSmith2010States,Chen2011}.
The latter technique has attracted particular attention in the
context of atomic magnetometry (see Ref.\
\cite{Koschorreck2010Decoupling} and references therein). We
note here that spin squeezing does not normally lead to
significant improvement of an optimized magnetometer measuring
quasistatic magnetic fields \cite{Auzinsh2004QND}; however, QND
techniques can extend the sensitivity to ac magnetic fields
\cite{Shah2010,Wasilewski2010}, and can improve the sensitivity
for cases involving nonexponential relaxation
\cite{Vasilakis2011}.

Let us briefly discuss quantum nondemolition measurements,
which minimally perturb the spin component being measured
\cite{Braginsky96,Geremia2006,Trail2010}. For example, consider
an ensemble of spin-1/2 atoms polarized in an unknown direction
perpendicular to the $y$-axis. In order to determine the angle
between the atomic polarization direction and the $x$-axis,
linearly polarized probe light propagating along the $x$-axis
can be employed. The polarization of the probe light will be
rotated due to circular birefringence depending on the
projection of the atomic spin along $\uv{x}$. If the probe
light is detuned from atomic resonance, the rate of atomic
transitions is low, so that the atomic polarization is not
destroyed. A continuous measurement produces a more and more
precise measurement of the spin projection along the $x$-axis,
indicating that the interaction with the probe light is
squeezing the atomic state. This means that the spin projection
in the orthogonal direction must be becoming more uncertain in
order to preserve the uncertainty relation. What is the
mechanism for this anti-squeezing? The linearly polarized probe
light can be thought of as being composed of a superposition of
left- and right-circularly polarized light, each of which
produce ac Stark shifts that mimic the effect of a magnetic
field directed along the $x$-axis. In the absence of noise,
these two fictitious magnetic fields nominally cancel, but the
effect of polarization noise (resulting from photon shot noise)
in the light beam is to produce an effective fluctuating
magnetic field along $\uv{x}$. The resulting fluctuating spin
precession causes a spread of the atomic state along the
$y$-axis, preserving the uncertainty relation. Note that this
process becomes more complicated for states with $F>1/2$ due to
the evolution of the internal degrees of freedom of the atoms,
as well as the coherence between the atoms (see Ref.\
\cite{Koschorreck2010Decoupling} and references therein). This
is discussed in more detail in Ref.\ \cite{Kurucz2010}; see
also Ref.\ \cite{OptMagBook}.

\section{Conclusion}

We have shown that the process of orientation-to-alignment
conversion, as induced by an electric field, is intimately
connected with spin squeezing. Following Kitagawa and Ueda
\cite{Masahito}, we quantified the amount of squeezing obtained
by this mechanism for a state of angular momentum $F$. While
the squeezing is improved for higher angular momentum, the
scalability of this approach is limited by the available values
of $F$ found in useable atomic and molecular states.
Alternative methods involving correlated atomic ensembles can
achieve much higher effective angular momenta; we briefly
discussed one such approach, using a quantum nondemolition
measurement.

\section*{Acknowledgments}

We are grateful to Dan M. Stamper-Kurn, Poul Jessen, Marcis
Auzinsh, Victor Acosta, Kasper Jensen, Irina Novikova, Morgan
Mitchell, Ivan Deutch, Max Zolotorev, Masahito Ueda, and
Michael Romalis for stimulating discussions. This research was
supported by Grant No.\ 2006220 from the United States-Israel
Binational Science Foundation (BSF), by NSF, and by the ONR
MURI program.

\bibliography{AtomLightInteractionsCurrent}

\appendix
\begin{widetext}
\section{Calculation of $F_+(t)$}

In order to determine the expression for the evolution of the
raising operator $F_+ (t)$ given in Eq.\ \eqref{Eq:upoperator},
we consider the matrix elements of $F_+ (t)$ between any two
eigenstates $|F,m \rangle$ of the Hamiltonian, so that
\begin{equation}
\begin{split}
\bra{F,m'}F_+ (t)\ket{F,m}
&=\bra{F,m'}e^{i\chi tF^{2}_z/\hbar^2}F_+ (0)e^{-i\chi tF^{2}_z/\hbar^2}\ket{F,m}, \\
&=e^{i\chi tm'^{2}}\bra{F,m'}F_+(0)\ket{F,m}e^{-i\chi tm^{2}}. \label{ren1}
\end{split}
\end{equation}
The only nonzero matrix elements of the raising operator are
those for which $m'=m+1$. For these matrix elements we have
\begin{equation}
\begin{split}
\bra{F,m+1}F_+(t)\ket{F,m}
&=e^{i\chi t (m+1)^2}\bra{F,m+1}F_+ (0)\ket{F,m} e^{-i\chi t m^2} \\
&=\bra{F,m+1}F_+(0)e^{i2\chi t(m+\frac{1}{2})}\ket{F,m}. \label{ren2}
\end{split}
\end{equation}
The energy eigenstates form a complete set, and so this
equation can be written in the operator form given in Eq.\
\eqref{Eq:upoperator}.

\section{Calculation of $\langle F_x (t) \rangle$}

Here we derive the expectation value \eqref{Eq:meanvalFx} of
the operator $F_x (t)$ [Eq.\ \eqref{Eq:Fx}] given the initial
state $\ket{F,F}_{\uv{x}}$ [Eq.\ \eqref{Eq:statecoupled}]:
\begin{equation}
\begin{split}
\abr{F_x(t)}
&= \bra{F,F}_{\uv{x}}\frac{1}{2}\sbr{F_+(0)e^{2i\chi t(F_z/\hbar+\frac{1}{2})}+e^{-2i\chi t(F_z/\hbar+\frac{1}{2})}F_-(t)}\ket{F,F}_{\uv{x}}\\
&= \re\sbr{\bra{F,F}_{\uv{x}}e^{-i2\chi t(F_z/\hbar+\frac{1}{2})}F_-(t)\ket{F,F}_{\uv{x}}}. \label{Eq:meanvalofxcomplexconj}
\end{split}
\end{equation}
We substitute Eqs.\ \eqref{Eq:statecoupled} and \eqref{Eq:downoperator} into Eq.\
\eqref{Eq:meanvalofxcomplexconj} and use the formula for the action of the $F_-(0)$ operator on the
eigenstates $\ket{Fm}$:
\begin{equation}
    F_-\ket{Fm}=\hbar\sqrt{F(F+1)-m(m-1)}\ket{F,m-1}.
\end{equation}
This results in
\begin{equation}
\abr{F_x(t)}=\hbar\sum^{2F}_{k,k'=0}\re\sbr{\bra{F,F-k'}2^{-2F}
  \binom{2F}{k'}^{\frac{1}{2}}e^{-i\mu(F_z/\hbar +\frac{1}{2})}\sqrt{F(F+1)-(F-k)(F-k-1)}
  \binom{2F}{k}^{\frac{1}{2}}\ket{F,F-k-1}},
\end{equation}
with $\mu=2\chi t$. Terms in the sum are nonzero only when
$F-k'=F-k-1$, i.e., $k'=k+1$. Thus we find
\begin{equation}
\begin{split}
\abr{F_x(t)}
&= \hbar\re\sbr{\bra{F,F-k-1}
2^{-2F}\sum^{2F}_{k=0}\sqrt{\binom{2F}{k+1}\binom{2F}{k}
}\sqrt{(2F-k)(k+1)}e^{-i\mu(F-k-1+\frac{1}{2})}\ket{F,F-k-1}}\\
&= \hbar\re\sbr{2^{-2F}\sum^{2F}_{k=0}\sqrt{\frac{2F!}{(2F-k-1)!(k+1)!}
 \frac{2F!}{(2F-k)!k!}(2F-k)(k+1)}e^{-i\mu(F-k-1+\frac{1}{2})}}\\
&= \hbar\re\sbr{2^{-2F}\sum^{2F}_{k=0}\sqrt{\frac{2F(2F-1)!}{(2F-k-1)!(k+1)k!}
 \frac{2F(2F-1)!}{(2F-k)(2F-k-1)!k!}(2F-k)(k+1)}e^{-i\frac{\mu}{2}(2F-2k-1)}}\\
&= \hbar\re\sbr{\frac{2F}{2^{2F}}\sum^{2F}_{k=0}\binom{2F-1}{k}e^{-i\frac{\mu}{2}(2F-k-1)}e^{i\frac{\mu}{2}k}}\\
&=\hbar\re\sbr{\frac{F}{2^{2F-1}}(e^{-i\frac{\mu}{2}}+e^{i\frac{\mu}{2}})^{2F-1}}\\
&=\frac{\hbar F}{2^{2F-1}}\prn{2\cos\frac{\mu}{2}}^{2F-1},
\end{split}
\end{equation}
where we have used the binomial formula
\begin{equation}
    \prn{x+y}^n=\sum_{j=0}^n\binom{n}{j}x^{n-j}y^j.
\end{equation}
Simplifying, we obtain Eq.\ \eqref{Eq:meanvalFx}.
\end{widetext}

\section{Squeezing due to orientation-to-alignment conversion in Cesium}

Here we apply the general expressions for squeezing due to
orientation-to-alignment conversion presented in the text for
the specific case of Cs atoms in the presence of a uniform
off-resonant light field. We assume stationary atoms (for
example, in a far-off-resonant optical trap) that are initially
in the $F=4$ hyperfine ground state and prepared in a stretched
state as assumed in the text. We apply a $z$-polarized light
field to the atoms detuned by $\GD$ from the D$_1$
$F=4\rightarrow F'=3$ or $F'=4$ transition, where $\GD$ is
greater than the natural width but is much smaller than the
splitting between hyperfine-structure levels, so that the
ground-state level shift is predominantly due to the ac Stark
effect arising from the interaction of the light with the
near-resonant transition.

The optimum value of the squeezing parameter for $F=4$ is
$\xi_{R}=0.6$, as can be found from Fig.\
\ref{fig:SqueezingFDpendence}. The time dependence of the
squeezing is given in terms of the quantum-beat frequency
$\chi$, which is found from the part of the Stark Hamiltonian
that is proportional to $F_z^2$ [see Eq.\
\eqref{Eq:IntHamiltonian}]. This term can be obtained by
writing
\begin{equation}\label{eq:acchi}
\hbar\chi = \bra{F,m=1}H_\text{eff}\ket{F,m=1} - \bra{F,m=0}H_\text{eff}\ket{F,m=0},
\end{equation}
where $H_\text{eff}$ is the effective ground-state Hamiltonian
describing ac-Stark shifts induced by mixing with the upper
state. For light tuned near the $F=4\rightarrow F'=4$
transition the second term is zero, and the first term is found
from second-order perturbation theory as
\begin{equation}
\bra{F,m=1}H_\text{eff}\ket{F,m=1} = \frac{\abs{\bra{F,m=1}d_zE\ket{F',m=1}}^2}{4\Delta}.
\end{equation}
Evaluating the dipole matrix element in terms of the reduced
matrix element (see, for example, Ref.\ \cite{ABRBook} for a
discussion of dipole matrix elements in the presence of hfs),
we find
\begin{equation}
\chi = \frac{\rme{J=1/2}{d}{J'=1/2}^2 E^2}{384\hbar^2\Delta};
\end{equation}
the value of the reduced dipole matrix element for the D$_1$
transition is $\rme{J=1/2}{d}{J'=1/2} = 3.2\,e a_0$. For light
tuned near the $F=4\rightarrow F'=3$ transition, both terms of
Eq.\ \eqref{eq:acchi} are nonzero, but the magnitude of $\chi$
works out to be the same.

From the formula \eqref{Eq:squeezingparameter} for the
squeezing parameter it can be shown that the optimal value for
$F=4$ occurs at the time when $\chi t = 0.036\times2\pi$.
\end{document}